\documentstyle[float,twocolumn,epsfig,amssymb,prb,aps]{revtex}
\begin{document}
\newfloat{figure}{ht}{aux}
%\initfloatingfigs
\draft
\twocolumn[\hsize\textwidth\columnwidth\hsize\csname
@twocolumnfalse\endcsname
\title{$S=1/2$ Chain-Boundary Excitations 
in the Haldane Phase of 1D $S=1$ Systems}
\author{E.~Polizzi, F.~Mila, and E.~S.~S\o rensen}
\address{Laboratoire de Physique Quantique IRSAMC, Universite Paul
Sabatier,\\ 118 Route de Narbonne, F-31062 Toulouse Cedex 4, France}
\date{\today}
\maketitle
\begin{abstract}
The $s=1/2$ chain-boundary excitations occurring in the Haldane
phase
of $s=1$ antiferromagnetic spin chains are investigated. The
bilinear-biquadratic hamiltonian is used to study these excitations
as a function of the strength of the biquadratic term, $\beta$,
between $-1\le\beta\le1$. 
At the AKLT point, $\beta=-1/3$ we show explicitly that these excitations are
localized at the boundaries of the chain on a length scale equal
to the correlation length $\xi=1/\ln 3$, and that the on-site
magnetization for the first site is $<S^z_1>=2/3$. 
Applying the density matrix
renormalization group we show that the chain-boundary
excitations remain localized at the boundaries for $-1\le\beta\le1$. 
As the two critical points $\beta=\pm1$ are approached the size of
the $s=1/2$ objects diverges and their amplitude vanishes.
\end{abstract}
%\pacs{75.10.-b, 75.10.Jm, 75.40.Mg}
\vskip1pc
]
%============================================================================
% BODY OF PAPER

%\section{Introduction}\label{sec:intro}

\begin{figure}
\begin{center}
\epsfig{file=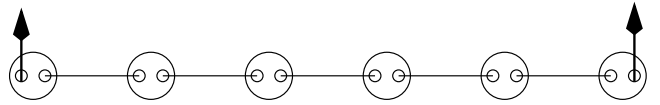,width=8.5cm}
\caption{Schematic representation of the VBS state for open boundary
conditions. Large circles
correspond to spin-1 sites and the smaller one to spin-1/2 states.
The $s=1/2$ excitations at the boundaries are indicated by the bold
arrows.}
\label{fig:vbs}
\end{center}
\end{figure}

Many different compounds exist which are believed to be
quasi one-dimensional $s=1$ antiferromagnetic systems. Among
these are CsNiCl$_3$, which
yielded the first experimental confirmation~\cite{expgap} of the Haldane gap~\cite{haldane},
as well as 
Ni(C$_2$H$_8$N$_2$)$_2$NO$_2$(ClO$_4$) (NENP)~\cite{renard87,expNENP}
which has been studied extensively and
Y$_2$BaNiO$_5$~\cite{spuche1,darriet,ditusa}, a compound similar
to the perovskites, which has garnered considerable attention recently due
to the relative ease with which it is carrier doped.
These compounds are thought to be in the so-called 
Haldane phase at low-temperatures.
This
phase is characterized by a disordered singlet ground-state with a gap to the
lowest lying excitations in the thermodynamic limit. 
In general, antiferromagnetically coupled linear $s=1$ systems
are believed to be in the Haldane phase if the hamiltonian describing
the interaction between the spins is sufficiently close to the standard
nearest neighbor Heisenberg model. In particular it is known that if
a biquadratic term linear in $\beta$ is added:
\begin{equation}
H=J\sum_{i=1}^{L-1}\left[{\bf S}_i\cdot{\bf S}_{i+1}-
\beta
({\bf S}_i\cdot{\bf S}_{i+1})^2
\right],
\label{eq:biham}
\end{equation}
the system remains in the Haldane phase for $-1<\beta<1$. 
At the Affleck-Kennedy-Lieb-Tasaki~\cite{AKLT} (AKLT) point,
$\beta=-\frac{1}{3}$, this model is partly solvable and the exact ground-state
corresponds to a valence bond solid (VBS) of adjacent dimers~\cite{AKLT}
(Fig.~\ref{fig:vbs})
and throughout the Haldane phase the ground-state is presumably closely
related to the VBS state. If each $s=1$ spin is viewed as two $s=1/2$
spin in the symmetric triplet state then the VBS state corresponds to
combining each of the two $s=1/2$ into a singlet with spins on the
adjacent sites.

As was noted by Kennedy~\cite{kennedy}, it is only when periodic boundary
conditions are imposed that the ground-state can be considered as a
non-degenerate singlet. If the chain is broken (i.e. open
boundary conditions) then the ground-state is actually four-fold
degenerate in the thermodynamic limit. 
One way to understand this four-fold degeneracy is to think of the
open chain as a periodic chain with one coupling removed. At the
AKLT point this corresponds to breaking one of the dimers, resulting
in two free $s=1/2$ objects and a ground-state that is four-fold
degenerate in the thermodynamic limit, as shown in Fig.~\ref{fig:vbs}. 
Hence, it was suggested~\cite{IanBert}
that such $s=1/2$ are real and experimentally observable.
For a finite open chain the $s=1/2$ excitations couple
and an exponentially
low-lying triplet, above the singlet ground-state, is found
in the Haldane gap~\cite{kennedy}.
In the Haldane phase
these $s=1/2$ objects can be
shown to be located at the boundaries of the open chain and we shall
refer to them as chain-boundary excitations. 

The chain-boundary excitations may seem like a curiosity, but it is
important to realize that the Haldane systems are gapped
and hence, in the presence of impurities,  the low-temperature
physics will be determined by the interaction between
these $s=1/2$ objects. In particular, there is considerable current
interest in disordered $s=1$
chains~\cite{rosskun,jolicoeur,orignac,spinimp}
which in many cases are treated in terms of interacting $s=1/2$ objects.
Hence, considerable effort has been devoted to 
the experimental verification of the above considerations.
Electron-spin-resonance (ESR) studies on Cu doped NENP~\cite{IanBert,hagiwara}
have confirmed the existence of these $s=1/2$ chain-boundary excitations.
Subsequent studies of NENP doped with non-magnetic impurities Zn, Cd and
Hg also showed evidence for $s=1/2$ excitations~\cite{glarum}. More
recent specific heat measurements~\cite{ramirez} on doped
Y$_2$BaNiO$_5$ has been interpreted as low-lying $s=1$ excitations,
but a more detailed analysis showed that an
interpretation in terms of $s=1/2$ excitations also is
possible~\cite{hallberg}.

In light of the above remarks 
it seems important to understand how the chain-boundary
excitations occur throughout the Haldane phase, a point we systematically
investigate in the present paper. In order to include
exact results we have studied the bilinear-biquadratic model,
Eq.~(\ref{eq:biham}), with the physical systems corresponding to $\beta=0$,
and we take this model as representative of the Haldane phase.
At the AKLT point the $s=1/2$ excitations can be studied using
the exact results for the VBS state as we detail below, facilitating
the analysis. Exponentially localized $s=1/2$ objects are shown
to occur at each end of an open chain.
Using the density matrix renormalization group (DMRG) we show
that these $s=1/2$ chain-boundary excitations remain well-defined
in the Haldane phase as $\beta$ is varied between $-1\le\beta\le 1$,
disappearing as the critical points $\beta=\pm1$ are approached.

%\section{Exact Results}

%Before turning to our exact results for the AKLT point we briefly
We now briefly
discuss the physics of the Hamiltonian Eq.~(\ref{eq:biham}). Two
critical points exist. The Uimin-Lai-Sutherland (ULS) point~\cite{uls},
described by a SU(3)$_{k=1}$ Wess-Zumino-Witten (WZW) model,
at $\beta=-1$, where the system displays a Berezinskii-Kosterlitz-Thouless
phase transition with an exponentially diverging correlation length
into a massless phase~\cite{itoi} for $\beta<-1$,
and the Takhtajan-Babujian (TB) point~\cite{tb},
described by a SU(2)$_{k=2}$ WZW model, at $\beta=1$ where
a second order phase transition to a dimerized phase occurs. 
At the ULS point the system has gapless modes at $k=0,\pm2\pi/3$
in contrast to the Heisenberg point, $\beta=0$, where the
Haldane gap occurs at $k=\pi$. Consequently, as the ULS point is approached,
a disorder point, which has been identified with the AKLT
point~\cite{scholwock}, occurs where the system develops short-range
incommensurate correlations. The associated Lifshitz point, where the
peak in the structure factor moves away from $k=\pi$, has been
numerically estimated to occur at
$\beta\simeq-0.438$~\cite{scholwock,bursill}.
At the TB point gapless modes occur at $k=0,\pi$ and hence, as $\beta$
approaches 1 the peak in the structure factor remains at $\pi$.

\begin{figure}
\begin{center}
\epsfig{file=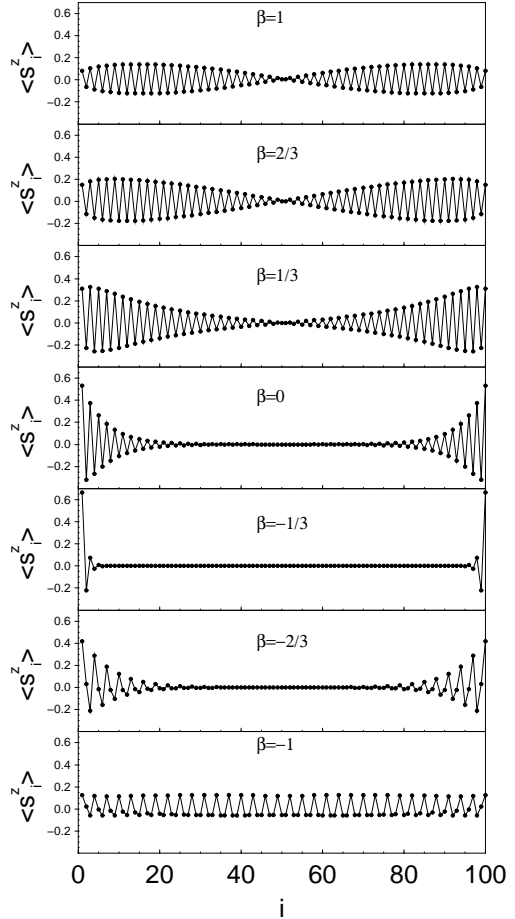,height=12.5cm}
\caption{The on-site magnetization $<S^z_i>$ along the
chain for different values of the biquadratic coupling,
$\beta$. In all cases $S^z_{T}=1$ was used.
}
\label{fig:szbeta}
\end{center}
\end{figure}

{\it Exact Results:} The VBS ground-state can be
written~\cite{AKLT,kolezhuk,arovas}
\begin{equation}
|\Psi_{\rm vbs}>=\prod_i g_i,\ 
g_i=\left(\matrix{{\frac{1}{\sqrt{2}}}|0>_i & -|+>_i\cr
|->_i & -\frac{1}{\sqrt{2}}|0>_i}\right).
%g_i=\left(\matrix{\sqrt{\frac{1}{3}}|0>_i & -\sqrt{\frac{2}{3}}|+>_i\cr
%\sqrt{\frac{2}{3}}|->_i & -\sqrt{\frac{1}{3}}|0>_i}\right).
\label{eq:gs}
\end{equation}
Here $|+>_i, |0>_i, |->_i$ corresponds to the three states of the
$s=1$ on site $i$. If periodic boundary conditions are considered
the trace should be taken. Using the above wave-function it is now
straightforward to evaluate $<S^z_1>_{\rm vbs}$, the magnetization
on the first site of the open chain, as a function of $L$. One finds
for the state with total magnetization $S^z_{T}=1$:
\begin{equation}
<S^z_1>_{\rm
vbs}(L)=\frac{\frac{2}{3}-2\cdot3^{-L}}{1-3^{-L}},\ L\ {\rm even}
\end{equation}
for even $L$, 
including all finite size corrections. 
As is evident from this
expression this amplitude rapidly approaches $2/3$ 
as $L\to\infty$. 
For odd L one finds
for the state with total magnetization $S^z_{T}=1$:
\begin{equation}
<S^z_1>_{\rm
vbs}(L)=\frac{\frac{2}{3}+2\cdot3^{-L}}{1+3^{-L}},\ L\ {\rm odd}.
\end{equation}
with the general expression for any site and any even $L$:
\begin{equation}
<S^z_i>_{\rm
vbs}(L)=(-1)^{i-1}\frac{\frac{2}{3}\left(\frac{1}{3}\right)^{i-1}
-2\cdot3^{i-1-L}}{1-3^{-L}}\ .
\label{eq:siz}
\end{equation}
Hence, in the thermodynamic limit one obtains the following result for
$<S^z_i>$ in the state $S^z_{T}=1$:
\begin{equation}
<S^z_i>_{\rm
vbs}=\frac{2}{3}\left(\frac{-1}{3}\right)^{i-1}
       =(-1)^{i-1}\frac{2}{3}e^{-(i-1)\ln 3}\ .
\label{eq:sztherm}
\end{equation}
This {\it explicitly} shows that the on-site magnetization is decreasing 
exponentially away from
the boundary of the chain with a length-scale 
equal to the bulk-correlation length, $\xi=1/\ln 3$~\cite{AKLT,arovas}.
Within the framework of a 
free boson
model~\cite{freeboson},
the same exponential decrease has been shown to occur at the Heisenberg
point~\cite{anis}, $\beta=0$, in accordance with numerical 
results~\cite{white,anis}. 
Hence
at the AKLT point the $s=1/2$ chain-boundary excitations are
well-defined objects localized at the two ends of the chain on a length
scale equal to the bulk correlation length. Presumably this picture
holds in most of the Haldane phase as we shall verify numerically
in the following.

%\section{DMRG Results}\label{sec:res}
{\it DMRG Results:} Away from the AKLT point the ground-state is not known and in order to
study the behavior of the chain-boundary excitations we use the density
matrix renormalization group (DMRG) method~\cite{white}. We refer the
reader to Ref.~\onlinecite{white} for a discussion of this method.
The calculations were performed using the total z-component 
of the
spin, $S^z_{T}$, as a quantum number and 
a number of states, $m$, between 60 and 100 was kept. The obtained 
results were extrapolated to the $m\to\infty,\ L\to\infty$ limit where needed. 

The bulk of our results are summarized in Fig.~\ref{fig:szbeta} where
$<S^z_i>$, the on-site magnetization, is plotted as a function of site
index, $i$, for 8 different values of $\beta$ between $-1\le\beta\le1$.
The calculations were in all cases performed in the $S^z_{T}=1$ state
with $m=100$. If the lowest-lying excitations are indeed determined
by the $s=1/2$ chain-boundary excitations, then selecting this state
corresponds to polarizing the two $s=1/2$ objects, and one would expect
a rather large signal at the ends of the chain. This is clearly the
case for most intermediate values of $\beta$ where well-defined
excitations are visible at the boundaries. 
Decreasing $\beta$ below the disorder point at $\beta=-1/3$ clearly
induces short-range incommensurate (IC) real-space correlations as the 
period-3
ground-state at $\beta=-1$ is approached, as can be seen in
Fig.~\ref{fig:szbeta} for $\beta=-2/3,\ -1$. Increasing $\beta$ beyond
approximately $\beta\sim0.2$ moves the maximum in the on-site
magnetization, $<S^z_i>$, away from the chain-boundaries, $i=1,L$. As
$\beta$ approached the critical point at $\beta=1$ the on-site
magnetization resembles more and more a double-doughnut shape which
presumably is characteristic of the dimerized state the system enters.

\begin{figure}
\begin{center}
\epsfig{file=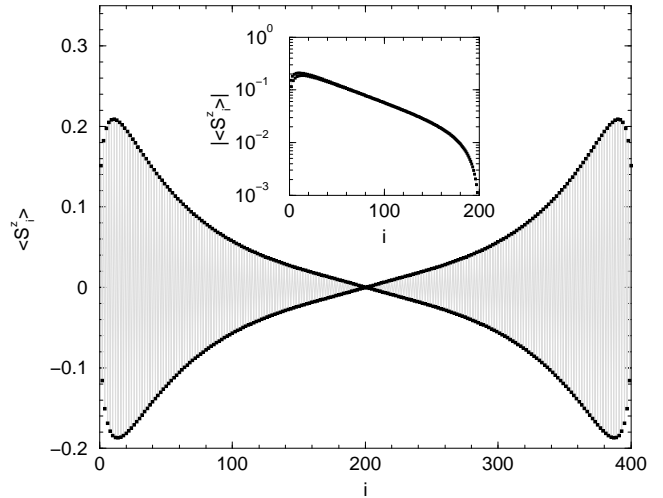,width=8.5cm}
\caption{The on-site magnetization in the $S^z_{T}=1$ state
for a 400
site chain with
$\beta=\frac{2}{3}$. The inset shows the same data for half the chain
on a log-scale.
The asymptotic exponential dependence is clearly visible. $m=100$ was used.
}
\label{fig:sz400}
\end{center}
\end{figure}

\begin{figure}
\begin{center}
\epsfig{file=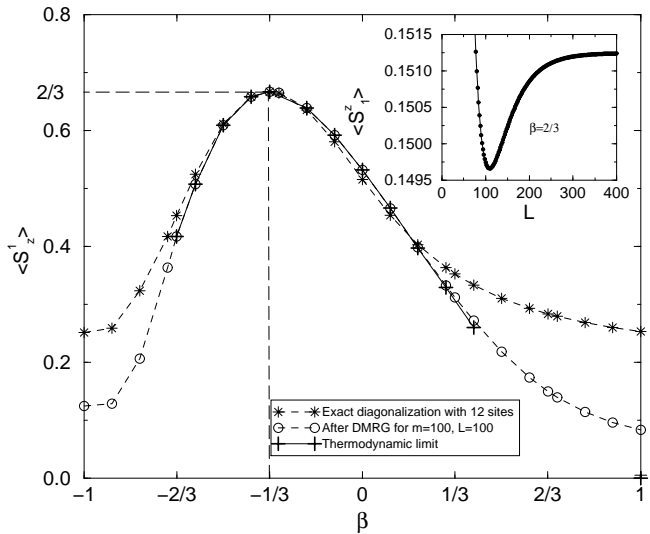,width=8.5cm}
\caption{The first site magnetization, $<S_1^z>$, as 
a function of $\beta$ between $-1<\beta<1$. The stars are
data for 12 sites ($\star$), the circles
data for $L=100,\ m=100$ ($\circ$), and the plusses extrapolations
in $m,L$ to the thermodynamic limit ({\bf +}).
The inset shows the $L$ dependence of $<S^z_1>$ for $\beta=\frac{2}{3}$
calculated with $m=100$.
}
\label{fig:sz1}
\end{center}
\end{figure}

Considering the results in Fig.~\ref{fig:szbeta} one may ask if the
chain-boundary excitations remain localized at the boundaries as the
critical points are approached. In order to obtain a partial answer to
this question we have re-examined the results $\beta=2/3$ considering
a system of size
$L=400$. Our results are shown in Fig.~\ref{fig:sz400}. An exponential
decay is now clearly visible as detailed in the inset. Assuming that
we can still identify the associated length scale with the bulk
correlation length we obtain $\xi=61(4)$. Hence, presumably the
$s=1/2$ objects remain localized at the boundaries until the critical
point $\beta=1$ is reached where the correlation length diverges. 

If the $s=1/2$ objects remain localized at the boundaries of the 
chain the first-site magnetization $<S^z_1>$ should remain finite
in the thermodynamic limit. If at some point the $s=1/2$ objects
de-localize and spread out over the entire chain we would expect
$<S^z_1>$ to go to 0 in the thermodynamic limit with non-trivial
finite-size corrections. As a measure of the presence of well-defined
chain-boundary excitations we have therefore numerically 
studied $<S^z_1>$ as a function of $\beta$. Two effects complicate 
the analysis: As $\beta\to\pm1$ the
correlation length rapidly diverges necessitating the use of larger
and larger system sizes as well as larger values of $m$, and secondly
the incommensurate correlations arising beyond the AKLT point. Our
results for several different values of $L,\ m$ are shown in Fig.~\ref{fig:sz1}.
The AKLT point constitutes an apparent maximum in this amplitude
with a value of $<S^z_1>=2/3$. Where possible, we have performed
extrapolations in both $L,m$ to obtain results representative of the
thermodynamic limit. These are shown as the solid line in
Fig.~\ref{fig:sz1}. The results are consistent with a non-zero value of
$<S^z_1>$ throughout the Haldane phase, vanishing as the two critical
points are approached at $\beta=\pm1$. The extrapolation to the
thermodynamic limit is further complicated due to the fact that
the $s=1/2$ objects strongly interact when $L\ll\xi$, leading to non-trivial
finite-size corrections to $<S^z_1>$. This is shown in the inset of
Fig.~\ref{fig:sz1} where $<S^z_1>$ is plotted as a function of $L$
for $\beta=2/3$. A clear minimum occurs at roughly 1-2 times the
estimated correlation length. We assume that this minimum corresponds
to a system size separating regions with strongly and weakly
interacting $s=1/2$ objects.

%\section{Conclusion}\label{sec:conclusion}
Summarizing, we have presented results that show that $s=1/2$
chain-boundary excitations occur throughout the Haldane phase 
as the biquadratic coupling is varied. These excitations are
localized at the ends of the chain with a maximum in their amplitude
of $<S^z_1>=2/3$ occurring  at the AKLT point, $\beta=-1/3$.
It would be interesting to compare our results for $<S^z_i>$,
proportional to the local susceptibility, with NMR results in addition
to the ESR results already available.

%
%\acknowledgments
%
We gratefully acknowledge helpful discussions with I.~Affleck, Th. Jolic\oe ur
and
K. Penc and thank IDRIS (Orsay) for an allocation of CPU time.

\end{document}